\documentclass[hidelinks]{article}
\usepackage[utf8]{inputenc}
\usepackage[left=3cm,right=3cm,top=3cm,bottom=3cm]{geometry}
\usepackage{pdfpages}
\usepackage{hyperref}
\usepackage{listings}
\usepackage{xcolor}
\usepackage{setspace} 
\usepackage{tocloft}
\usepackage{amsmath}
\usepackage{chngcntr}
\usepackage[toc,page]{appendix}
\usepackage[T1]{fontenc}
\usepackage[nottoc]{tocbibind}
\usepackage[compact]{titlesec}
\usepackage[utf8]{inputenc}
\usepackage{natbib}
\usepackage{amsmath}
\usepackage{amssymb}
\usepackage{url}

\usepackage{multirow}
\usepackage{setspace}
\usepackage{array}
\usepackage{float}

\newcolumntype{L}[1]{>{\raggedright\let\newline\\\arraybackslash\hspace{0pt}}m{#1}}
\newcolumntype{C}[1]{>{\centering\let\newline\\\arraybackslash\hspace{0pt}}m{#1}}
\newcolumntype{R}[1]{>{\raggedleft\let\newline\\\arraybackslash\hspace{0pt}}m{#1}}

\usepackage{subfig}
\usepackage{bm}
\usepackage{multirow}
\usepackage{tikz}
\usetikzlibrary{shapes,arrows}
\usepackage{blkarray}
\usepackage[affil-it]{authblk}
\usepackage{listings}
\usepackage{siunitx}
\usepackage{lineno}

\providecommand{\keywords}[1]{\textbf{\textit{Keywords ---}} #1}

\title{On the association between COVID-19 vaccination levels and incidence and lethality rates at a regional scale in Spain}

\author[1*]{Álvaro Briz-Redón}
\author[2*]{Ángel Serrano-Aroca}

\affil[1]{Department of Statistics and Operations Research, University of Valencia, Spain}
\affil[2]{Centro de Investigación Traslacional San Alberto Magno, Universidad Católica de Valencia San Vicente Mártir, Spain}

\affil[*]{\normalfont Corresponding author}

\date{}

\begin{document}

\maketitle

\newpage

\begin{abstract}

The severe acute respiratory syndrome coronavirus 2 (SARS-CoV-2), which causes the coronavirus disease 2019 (COVID-19), has led to the deepest global health and economic crisis of the current century. This dramatic situation has forced the public health authorities and pharmaceutical companies to develop anti-COVID-19 vaccines in record time. Currently, almost 80\% of the population are vaccinated with the required number of doses in Spain. Thus, in this paper, COVID-19 incidence and lethality rates are analyzed through a segmented spatio-temporal regression model that allows studying if there is an association between a certain vaccination level and a change (in mean) in either the incidence or the lethality rates. Spatial dependency is included by considering the Besag-York-Mollié model, whereas natural cubic splines are used for capturing the temporal structure of the data. Lagged effects between the exposure and the outcome are also taken into account. The results suggest that COVID-19 vaccination has not allowed yet (as of September 2021) to observe a consistent reduction in incidence levels at a regional scale in Spain. In contrast, the lethality rates have displayed a declining tendency which has associated with vaccination levels above 50\%.

\end{abstract}


\newpage

\section{Introduction}

The coronavirus disease 2019 (COVID-19) pandemic has already claimed the lives of over 4.8 million people and more than 235 million confirmed cases worldwide \citep{WHO1}. The global health and economic crisis caused by this new coronavirus has highlighted the major role that vaccination could play in controlling the incidence, hospitalization and lethality caused by this disease with more than 50 novel vaccine candidates already in clinical trials \citep{gallagher2021indirect}. Currently there are more than 6000 million vaccine doses administered worldwide \citep{WHO1}. In Spain, there is already more than 70 million doses administered and more than 35 million vaccinated people with the required number of doses in a population of about 47 million people, which represents almost 80\% of the population already fully vaccinated \citep{Sanidad}. Four vaccines approved by the European Medicines Agency (EMA) that have shown variable protection against severe acute respiratory syndrome-Coronavirus-2 (SARS-CoV-2) variants have been administered in Spain so far: the lipid nanoparticle (LNP)-formulated mRNA COVID-19 vaccines BNT162b2 (Pfizer/BioNTech) \citep{sahin2020covid}, the mRNA-1273 (Moderna) \citep{jackson2020mrna},  the adenovirus (Ad)-based vaccines ChAdOx1 nCoV-19 (University of Oxford/AstraZeneca) \citep{mercado2020single} and the Ad26.COV2.S (Johnson \& Johnson/Janssen) \citep{arashkia2021severe}. However, there are mixed opinions about vaccine effectiveness (VE) considering that this virus can be easily transmitted through aerosols in poor ventilated places \citep{klompas2020airborne,prather2020airborne}, and also because of the presence of asymptomatic carriers \citep{wilmes2021sars}. The diverse mechanisms of action of the new vaccines administered are a subject of debate too \citep{leshem2021population}. Thus, vaccines can act with a first mechanism, also known as sterilising immunity, to block infection occurring entirely and, in this case, people cannot transmit the virus, or they can act with a mechanism consisting of stopping the progression of symptoms after infection occurs \citep{gallagher2021indirect}. However, in this second case, people can still transmit the infection to others, and the coronavirus can continue mutating and creating harmful new variants, which render this fight even more difficult. Currently, there are four variants of concern, the Alpha, Beta, Gamma and Delta variants, which were originally identified in the UK, South Africa, Brazil and India, respectively, as well as with an increasing number of other recently identified SARS-CoV-2 variants \citep{tao2021biological}. Particularly, the rapid spread of the Delta variant is a great risk to global public health \citep{farinholt2021transmission}, as it is highly transmissible and contains mutations that confer partial immune escape \citep{riemersma2021shedding}. Indeed, it has been shown that vaccination reduces transmission of Delta, but by less than the Alpha variant \citep{eyre2021impact}. In this context, it is not clear yet if the administered vaccines could provide protection against all coexisting variants and other variants expected to emerge in the near future. Furthermore, the durability of responses after COVID-19 is limited and new doses are required to keep the immunity strong against this pathogen \citep{widge2021durability}. Thus, in this study, we have performed a spatio-temporal analysis for exploring the large-scale effect of vaccination on COVID-19 incidence and lethality in Spain. The main goal is to analyze if the vaccination levels have already shown an association with a change in either the incidence or the lethality rates.


\section{Data}

All the data used to conduct this study was downloaded from the public repository available in https://github.com/montera34/escovid19data. This repository is maintained by volunteers that collect data from multiple sources, including the Institute of Health Carlos III (ISCIII) from Spain. The data is provided at different levels of spatial aggregation and on a daily basis. In the present study, data is analyzed at a regional level according to the division of the country in Autonomous Communities, although only those located in the Iberian Peninsula have been considered. The study period spans from 4 January 2021 (first day for which the datasets provide vaccination values) to 30 September 2021. Thus, three spatio-temporal variables have been constructed for each region $i$ and date $t$ within the study period: COVID-19 daily incidence rate ($I_{i,t}$), COVID-19 daily lethality rate ($L_{i,t}$), and COVID-19 daily vaccination level ($V_{i,t}$). Specifically, $I_{i,t}$, $L_{i,t}$, and $V_{i,t}$ have been computed as follows:
$$I_{i,t}=100000\times\frac{\text{Number of new cases detected in region} \hspace{0.1cm} i \hspace{0.1cm}  \text{on date} \hspace{0.1cm} t}{\text{Population of region} \hspace{0.1cm} i}$$
$$L_{i,t}=100\times\frac{\text{Cumulative number of deaths recorded in region} \hspace{0.1cm} i \hspace{0.1cm} \text{on date} \hspace{0.1cm} t}{\text{Cumulative number of cases detected in region} \hspace{0.1cm} i \hspace{0.1cm} \text{on date} \hspace{0.1cm} t}$$
$$V_{i,t}=100\times\frac{\text{Number of residents fully vaccinated in region} \hspace{0.1cm} i \hspace{0.1cm} \text{on date} \hspace{0.1cm} t}{\text{Population of region} \hspace{0.1cm} i}$$

\begin{figure}[H]
  \centering
  \subfloat[]{\includegraphics[width=5cm,angle=-90]{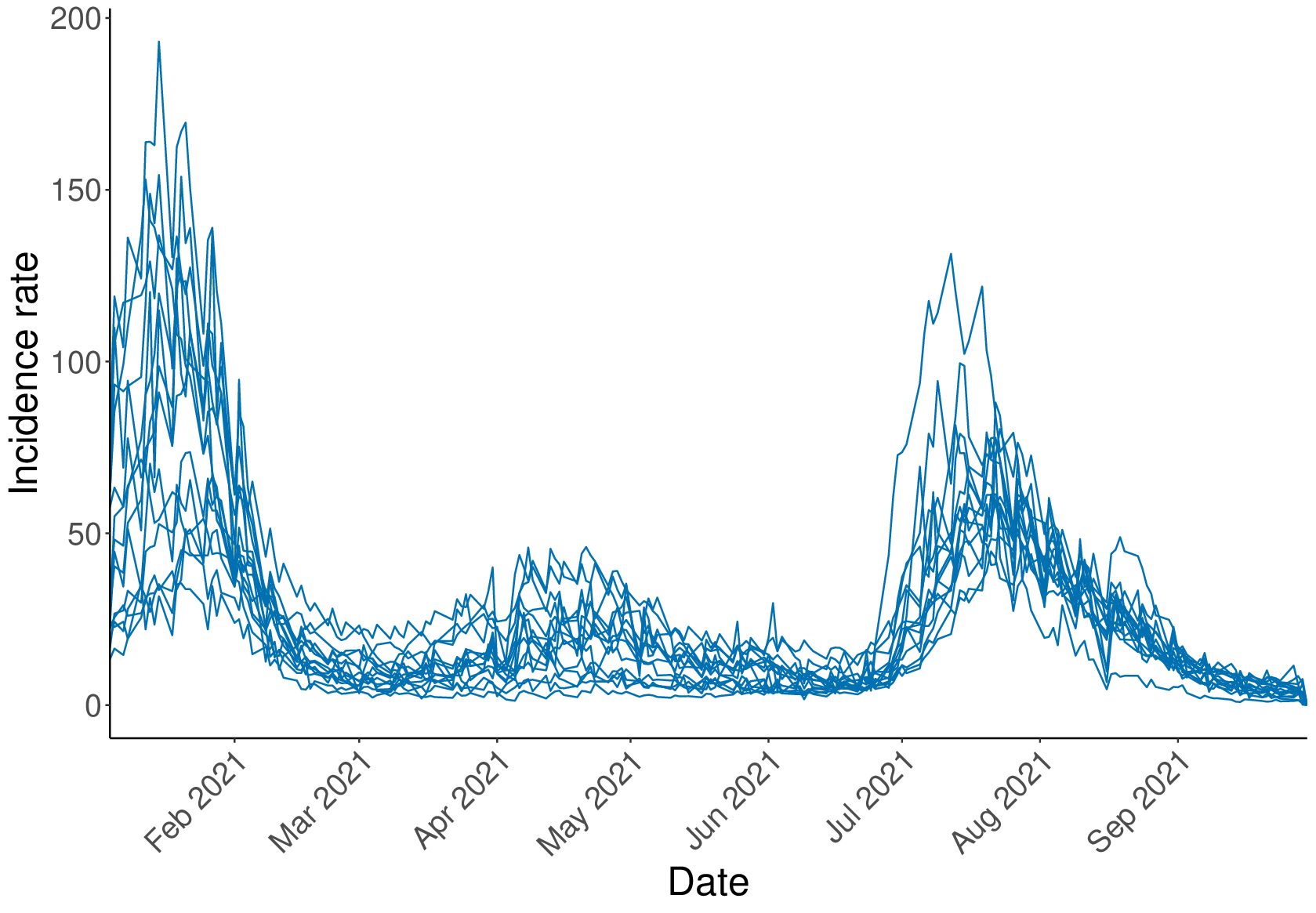}\label{fig:exploratory_a}}
  \subfloat[]{\includegraphics[width=5cm,angle=-90]{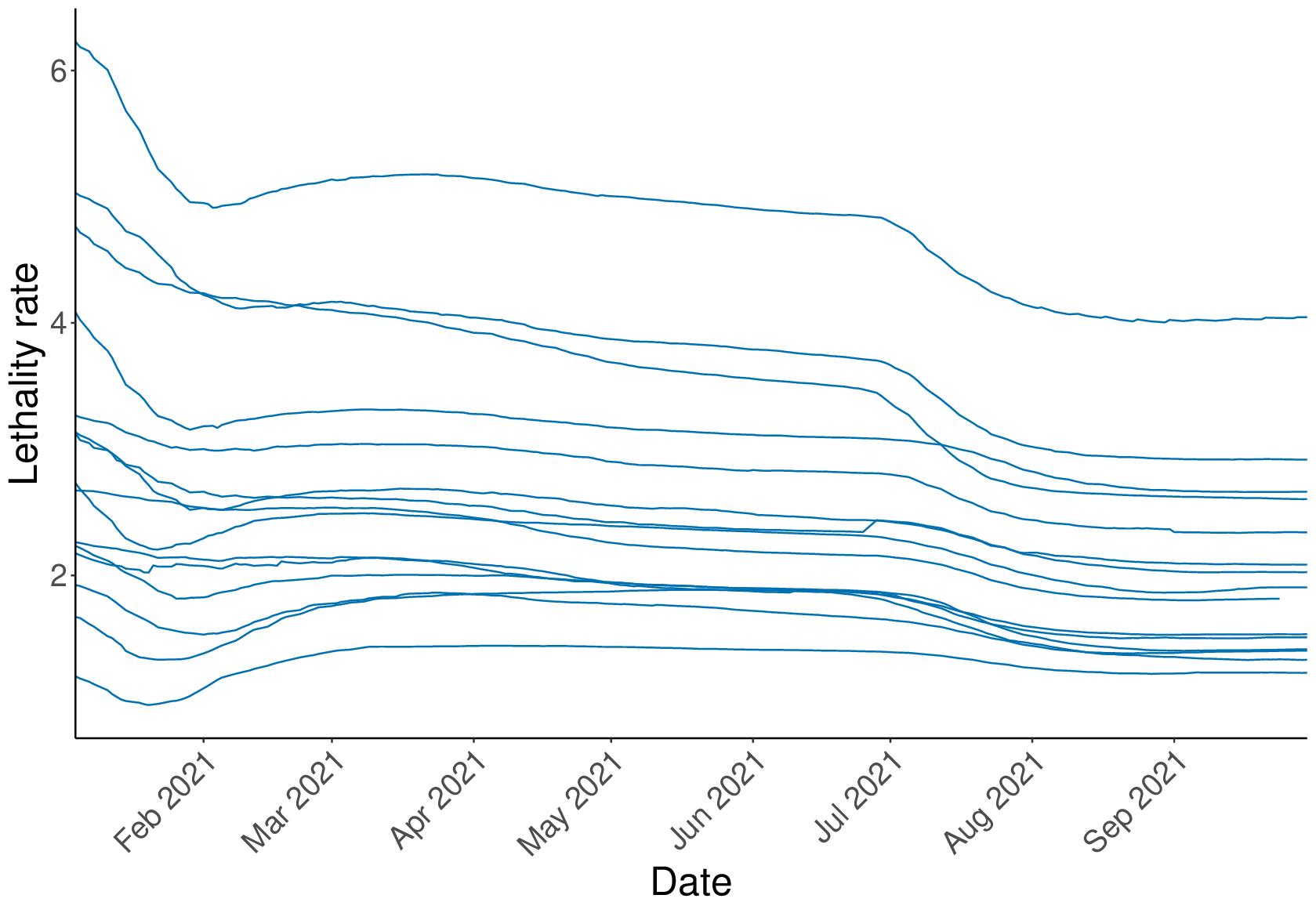}\label{fig:exploratory_b}}
 \caption{Evolution of incidence (a) and lethality (b) rates for the regions of Spain considered for the analysis during the period January 2021-September 2021}
 \label{fig:exploratory} 
\end{figure}

\begin{figure}[H]
  \centering
  \includegraphics[width=0.6\linewidth,angle=-90]{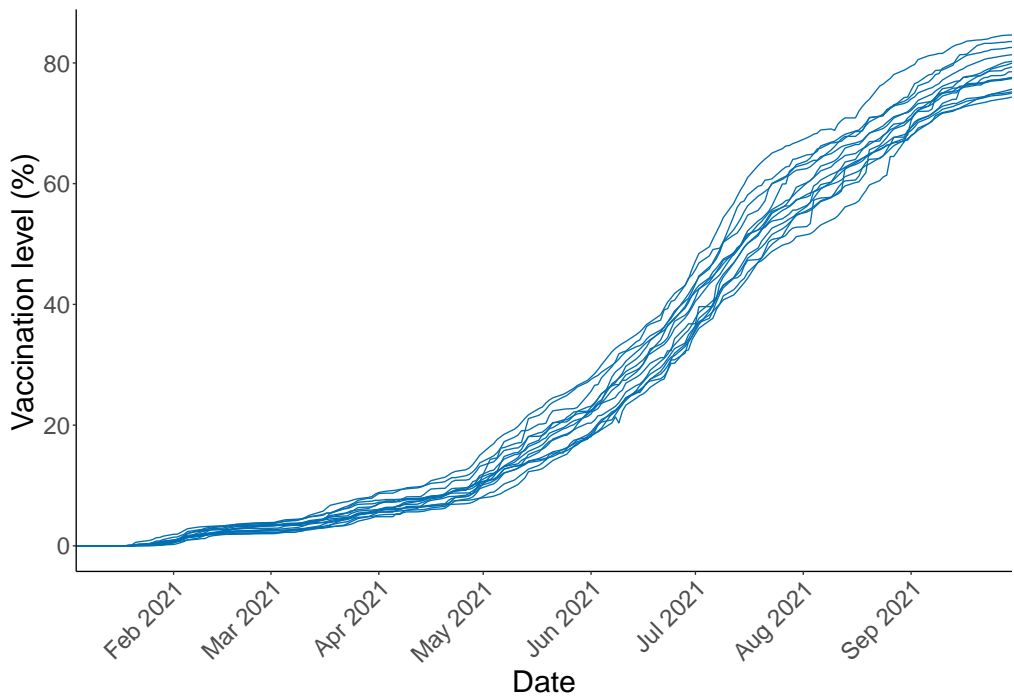}
   \caption{Evolution of (complete) vaccination levels for the regions of Spain considered for the analysis during the period January 2021-September 2021}
 \label{fig:vaccine} 
\end{figure}

The definition considered for the lethality corresponds to the case fatality ratio (CFR), which is widely used for measuring the severity of a disease given its simplicity \citep{ghani2005methods}, although other alternatives are also available \citep{kim2021estimation}. Incidence and lethality rates for the regions of Spain during the study period are shown in Figure \ref{fig:exploratory}. Vaccination levels since the beginning of 2021 for the same set of regions are shown in Figure \ref{fig:vaccine}.

\section{Methodology}

\subsection{Model description}

A change point represents an abrupt variation in time series data, possibly as a consequence of a transition between states \citep{aminikhanghahi2017survey}. More precisely, if $\{y_{t}\}_{t=1}^{n}$ is a time series, change point detection can be seen as a hypothesis test problem, considering a null hypothesis of the form ``No change in the time series occurs'', and an alternative hypothesis that states that there exists a time, $\tau \in \{1,...,n\}$, such as the statistical properties of $\{y_{1},...,y_{\tau}\}$ and $\{y_{\tau+1},...,y_{n}\}$ differ in some way \citep{killick2014changepoint}.

The basis behind change point analysis can be extended to account for the effect of a temporally-varying covariate potentially associated with a variation in the values of a time series of interest. This leads to models with change points in the covariates, which can be adapted to multiple contexts, including survival analyses \citep{lee2020testing} and exposure-outcome epidemiological studies \citep{sarnaglia2021estimating}. In general, segmented regression models \citep{muggeo2003estimating} are the natural framework for dealing with this type of covariate effects.

In this study, a segmented spatio-temporal regression is employed for analyzing COVID-19 incidence and lethality rates at the regional level in Spain. Specifically, $y_{i,t}$ denotes indistinctly the COVID-19 incidence or lethality in region $i$ ($i=1,...,15$) on date $t$ ($t=1,...,192$). The (complete) vaccination level in region $i$ on date $t$, denoted by $V_{i,t}$, is the only covariate considered for the analysis. Hence, the purpose of the model is to detect if the mean value of $y_{i,t}$ depends on the condition $V_{i,t}>c$ (or, equivalently, $V_{i,t}\leq c$). In other words, we are interested in determining if there is an association between a certain vaccination level, $c$, and a change (in mean) in either the incidence or the lethality rates.  

Thus, $y_{i,t}$ is specified according to the following spatio-temporal structure
\begin{equation}
y_{i,t}=\alpha+\beta \cdot I(V_{i,t-\mathrm{lag}}>c)+u_{i}+v_{i}+\mathrm{ns}(t,\mathrm{df})    
\label{eq:main_model}
\end{equation}
where $I(V_{i,t-\mathrm{lag}}>c)$, considering a temporal lag of 0, 7, or 14 days, is an indicator function that is equal to 1 if the vaccination level exceeds some given threshold, $c$, and 0 otherwise. The use of a temporal lag is convenient in the context of epidemiological studies as there can exist a delayed association between the exposure variable (vaccination level) and the outcome (either incidence or lethality rate) under study \citep{bhaskaran2013time}. Parameter $\alpha$ denotes the intercept of the model and $\beta$ the effect that a vaccination level above $c$ has on $y_{i,t}$. For studying the impact of vaccination levels on both regional incidence and lethality rates, the value of $c$ has been ranged from 10\% to 80\%, in intervals of 10 percentage points. The terms $u_{i}$ and $v_{i}$ represent, respectively, the structured and unstructured spatial random effect of the model, which are based on neighborhood relationships between regions (two regions have been considered as neighbors if they share a geographical boundary). Specifically, the Besag–York–Molliè (BYM) model has been considered \citep{besag1991bayesian}, which establishes, under a Bayesian framework, that the conditional distribution of the spatially-structured effect on area $i$, $u_{i}$, is
$$u_{i}|u_{j \neq i} \sim Normal\bigg(\sum_{j \neq i=1}^{n} w_{ij}u_{j},\frac{\sigma^2_{u}}{N_{i}}\bigg)$$
where $N_{i}$ is the number of neighbors for area $i$, $w_{ij}$ is the ($i$,$j$) element of the row-normalized neighborhood matrix ($w_{ij}=1/N_i$ if regions $i$ and $j$ are neighbors, and 0 otherwise), and $\sigma^2_{u}$ represents the variance of the random effect. The spatially-unstructured effect, denoted by $v_{i}$, follows a Gaussian distribution, $v_{i} \sim Normal(0,\sigma^2_{v})$, where $\sigma^2_{v}$ is the variance of this effect. Finally, $\mathrm{ns}(t,\mathrm{df})$ denotes a natural cubic spline on the temporal component with $\mathrm{df}$ degrees of freedom \citep{hastie2017statistical}. Natural splines are suitable for capturing trends and seasonal patterns in the data \citep{bhaskaran2013time}.

A modification of the model presented by Equation \ref{eq:main_model} was also tested by allowing the slope parameter ($\beta$) to vary regionally
\begin{equation}
y_{i,t}=\alpha+(\beta+\gamma_{i}+\delta_{i}) I(V_{i,t}>c)+u_{i}+v_{i}+\mathrm{ns}(t,\mathrm{df})
\label{eq:main_model_random_slope}
\end{equation}
where $\gamma_{i}$ and $\delta_{i}$ follow the same structure of the BYM model described above for $u_i$ and $v_i$. Thus, the addition of these parameters enables us to account for the possibility that the effect of a vaccination level above $c$ has differed across regions.

All the models have been fitted with the \textsf{INLA} R package \citep{INLA2}, which makes use of the Integrated Nested Laplace Approximation proposed by \cite{INLA1}. Non-informative priors for the fixed-effects parameters and the default Gamma-distributed priors for the precision of the random effects were employed.

\subsection{Software}

The R programming language \citep{teamR} has been used in our analysis. In particular, the R packages \textsf{ggplot2} \citep{ggplot2}, \textsf{rgdal} \citep{rgdal}, \textsf{rgeos} \citep{rgeos}, \textsf{spdep} \citep{bivand2008applied}, and \textsf{splines} \citep{splines} have also been used. 

\section{Results}

\begin{figure}[H]
  \centering
  \subfloat[]{\includegraphics[width=5cm,angle=-90]{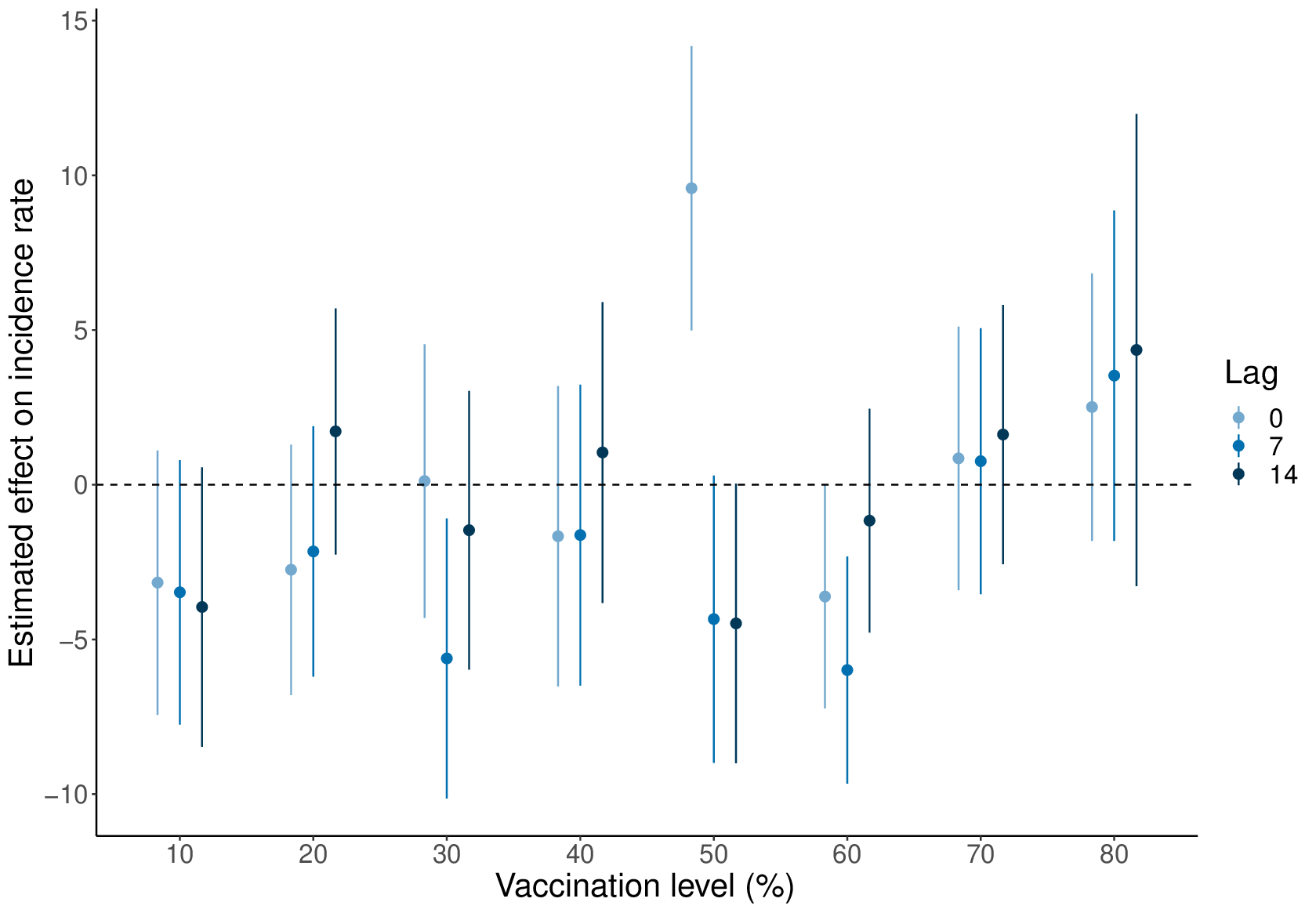}\label{fig:results_beta_a}}\hspace{0.1cm}
  \subfloat[]{\includegraphics[width=5cm,angle=-90]{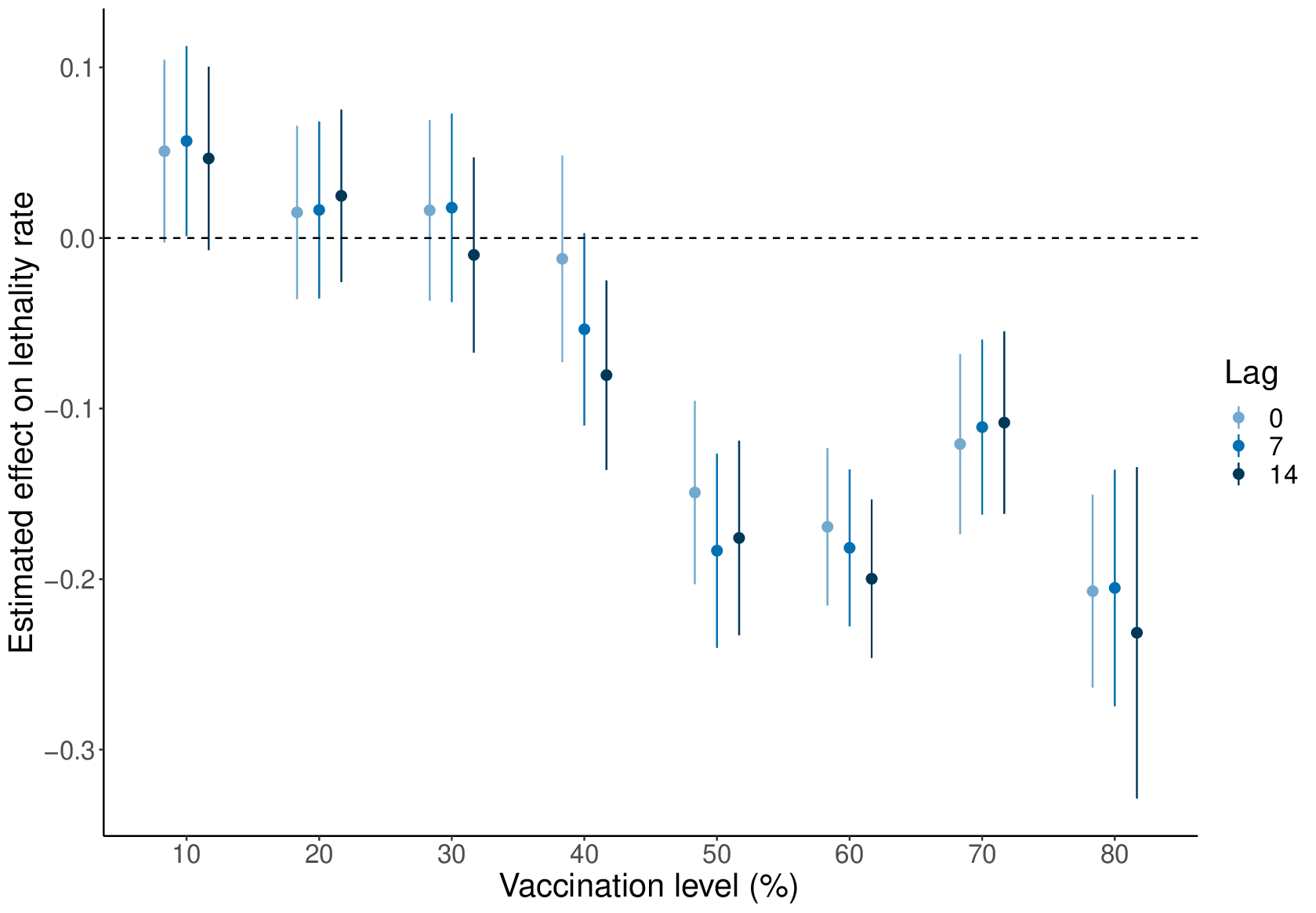}\label{fig:results_beta_b}}
 \caption{Estimated effect (with the 95\% credibility interval) of a certain vaccination level on incidence (a) and lethality (b) rates. These estimates correspond to $\beta$ in Equation \ref{eq:main_model}}
 \label{fig:results_beta} 
\end{figure}

Figure \ref{fig:results_beta} summarizes the results corresponding to fitting the model with change points in the vaccination levels represented by Equation \ref{eq:main_model} (natural cubic splines with 16 and 10 degrees of freedom were finally chosen for modeling incidence and lethality rates, respectively, as they yielded the best results according to the DIC proposed by \cite{spiegelhalter2002bayesian}). Specifically, the estimates of parameter $\beta$ are shown considering the incidence (Figure \ref{fig:results_beta_a}) and the lethality rates (Figure \ref{fig:results_beta_b}) as the response variable of the model. The 95\% credibility intervals corresponding to each estimate are also displayed. As mentioned earlier, temporal lags of 0, 7, and 14 days are tested to account for delayed exposure-outcome associations. In the case of the incidence rates, most of the $\beta$ estimates are not different from 0 with 95\% credibility, except for the estimates corresponding to $c=30\%$ ($\mathrm{lag}=7$), $c=50\%$ ($\mathrm{lag}=0$), and $c=60\%$ ($\mathrm{lag}=0$ and $\mathrm{lag}=7$). The first of these associations must be a consequence of a wave of COVID-19 that Spain suffered at mid-July 2021, when vaccination levels were around 50\%. Despite these results, the associations between vaccination levels and incidence rates for the study period, considering both the modeling approach chosen and the scale of the geographical units under analysis, do not show a clear trend. Indeed, the strong association for $c=50\%$ with $\mathrm{lag}=0$ disappears when a temporal lag of 7 or 14 days is considered, even though the impact of varying the temporal lag is generally minor.

In contrast, the estimates of the $\beta$ parameter for the models describing lethality rates suggest that lethality rates are decreasing progressively. Particularly, when vaccination reached the 50\% of the population, the estimates of $\beta$ are negative with 95\% credibility, which indicates a change point in the time series of lethality rates at this stage of the vaccination process. This should not be interpreted as a cause-effect relationship, but points out at which percentage of vaccination level the lethality rates of the regions of Spain might have started to display some reduction.

\begin{figure}[H]
  \centering
  \subfloat[]{\includegraphics[width=5cm,angle=-90]{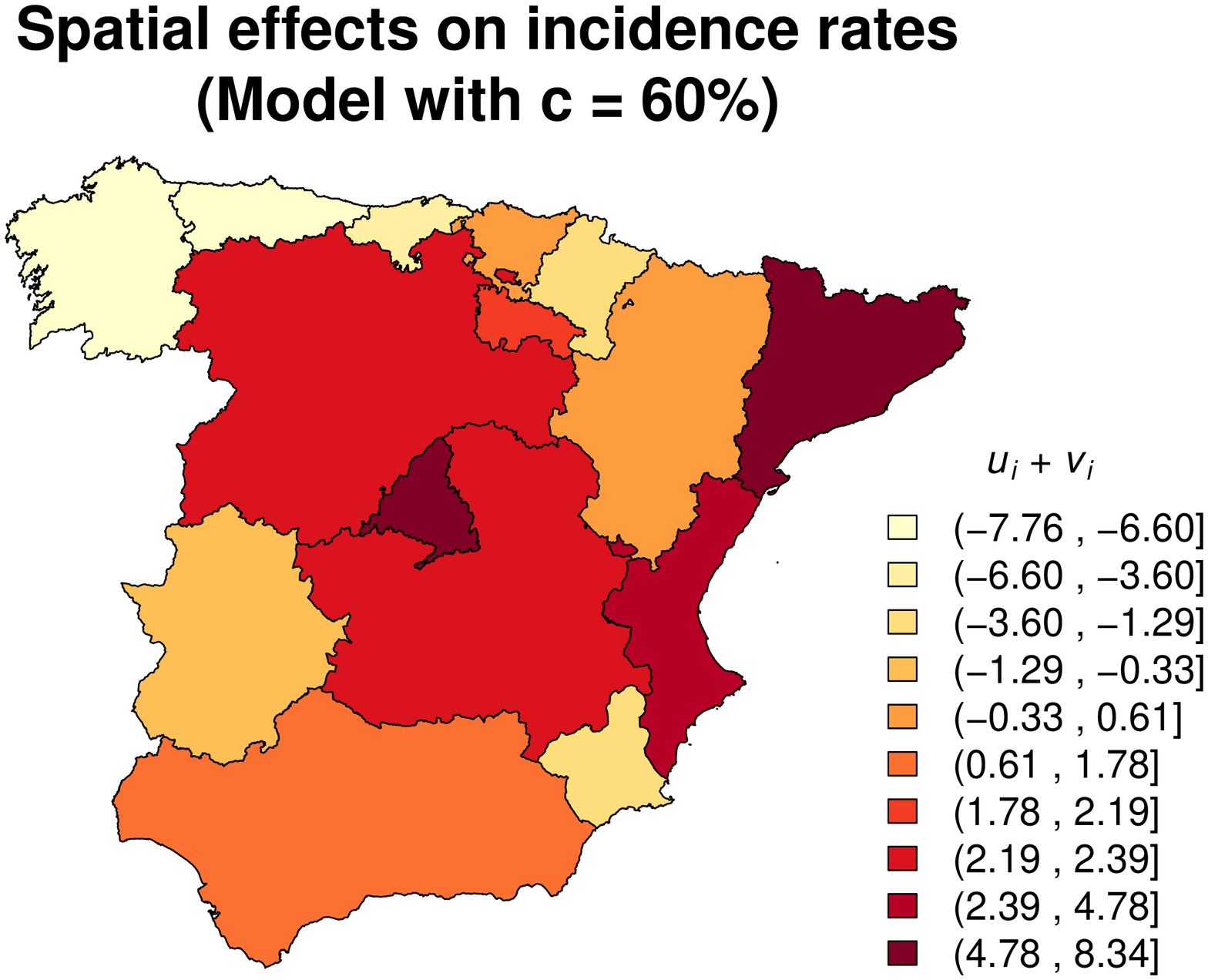}\label{fig:results_spatial_effects_a}}\hspace{0.1cm}
  \subfloat[]{\includegraphics[width=5cm,angle=-90]{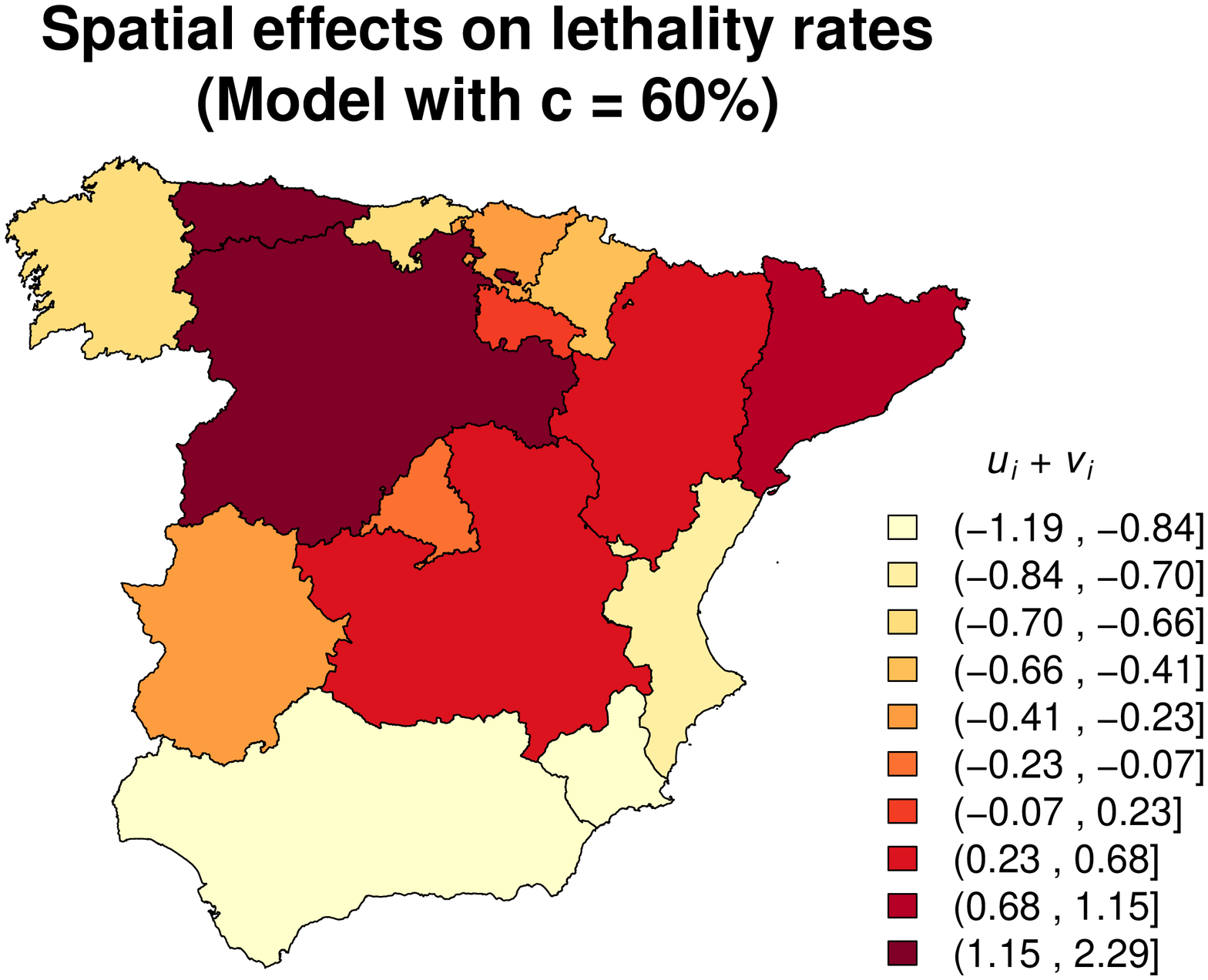}\label{fig:results_spatial_effects_b}}
 \caption{Estimated spatial effects at the region level for incidence (a) and lethality (b) rates, considering $\mathrm{lag}=7$. These estimates correspond to $u_i+v_i$ in Equation \ref{eq:main_model}}
 \label{fig:results_spatial_effects} 
\end{figure}

Furthermore, Figure \ref{fig:results_spatial_effects} shows the estimates of the spatial effects ($u_i+v_i$) on incidence (Figure \ref{fig:results_spatial_effects_a}) and lethality rates (Figure \ref{fig:results_spatial_effects_b}) for $c=60\%$. The results are nearly the same for similar vaccination levels. It can be appreciated that there exists a marked spatial variation in terms of both incidence and lethality rates. In particular, some of the more densely populated regions of Spain have experienced the highest incidence risk (Madrid, Cataluña, and the Comunidad Valenciana). However, lethality risk has been higher in some regions such as Asturias and Castilla y León, possibly as a consequence of the presence of more population in the older age ranges. The computation of the proportion of variance explained by each spatial effect \citep{blangiardo2015spatial} has revealed that the unstructured component, $v_i$, captures around 99\% and 95\% of the spatial variability of the incidence and lethality rates, respectively, at the regional level studied. This suggests that spatial proximity is not sufficient to explain the differences in incidence/lethality displayed by the considered regions of Spain during the last months of the COVID-19 pandemic.

\begin{figure}[H]
  \centering
  \subfloat[]{\includegraphics[width=5cm,angle=-90]{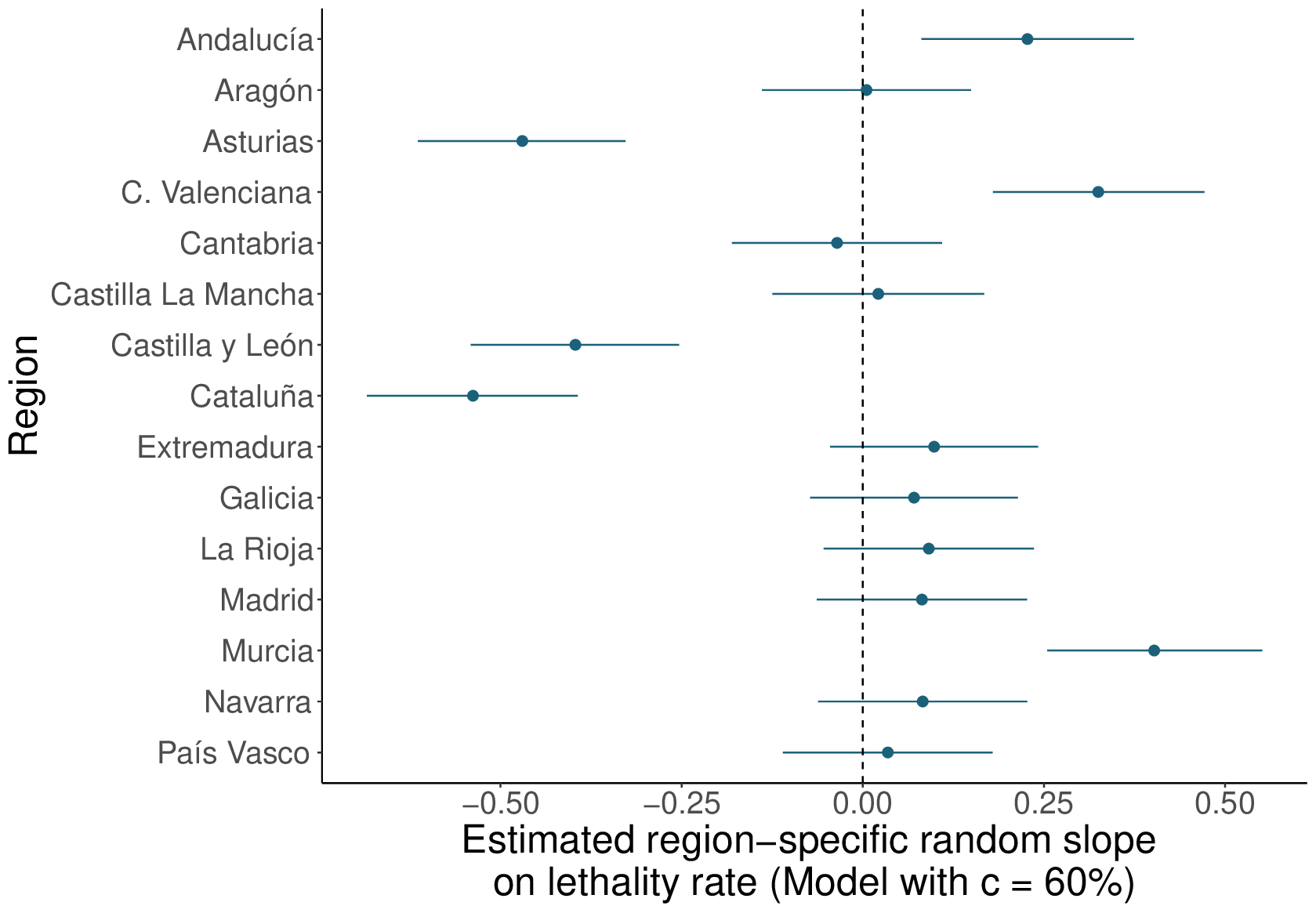}\label{fig:results_let_random_slope_a}}\hspace{0.1cm}
  \subfloat[]{\includegraphics[width=5cm,angle=-90]{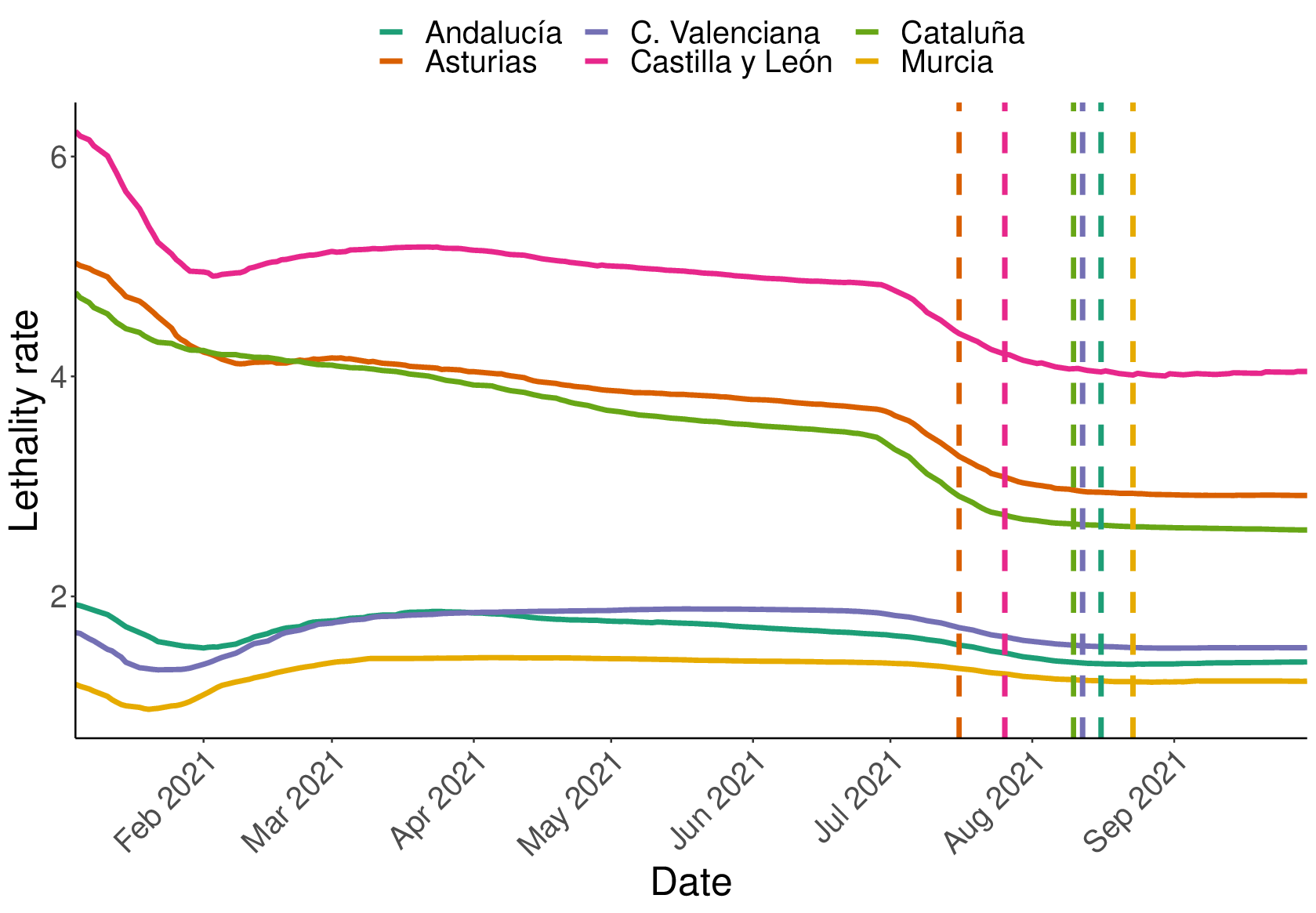}\label{fig:results_let_random_slope_b}}
 \caption{Estimated region-specific random slopes (with the 95\% credibility intervals) (a) considering the modeling of lethality rates (for $c=60\%$ and $\mathrm{lag}=7$). These estimates correspond to $\gamma_i+\delta_i$ in Equation \ref{eq:main_model_random_slope}. In (b), the evolution of lethality rates for six specific regions is shown. Each dashed vertical line represents the date at which vaccination level reached 60\% for the corresponding region}
 \label{fig:results_let_random_slope} 
\end{figure}

Finally, Figure \ref{fig:results_let_random_slope} summarizes the results that follow from fitting Equation \ref{eq:main_model_random_slope} to allow for regionally-varying covariate effects. Regarding this modeling approach, we have only focused on lethality rates, as no clear association has been found for incidence rates with the model represented by Equation \ref{eq:main_model}. Figure \ref{fig:results_let_random_slope_a} shows the estimates of $\gamma_i+\delta_i$ for each of the regions under analysis, considering $c=60\%$ and $\mathrm{lag=7}$. There are marked differences, as for six of the regions the estimates are different from 0 with 95\% credibility. Figure \ref{fig:results_let_random_slope_b} displays the evolution of the lethality rates for these regions, indicating the date at which vaccination levels reached 60\%. It can be observed than lethality rates dropped remarkably in Asturias, Castilla y León, and Cataluña at the time this level of vaccination was reached. Again, it must be noted that cause-effects relationships cannot be derived from this analysis, but only exposure-outcome associations.

\section{Discussion and conclusions}

According to a study performed in Israel, COVID-19 vaccination is highly effective in preventing symptomatic and asymptomatic SARS-CoV-2 infections and COVID-19-related hospitalisations, and deaths \citep{haas2021impact}. The studies of VE performed in Spain, where around 80\% of the population has already been vaccinated, have provided encouraging results so far. Thus, even though the vaccination campaign in Spain was carried out at a much slower speed than Israel, COVID-19 vaccination using mRNA vaccines in Spain has shown to be very effective in preventing infections, and COVID-19 hospitalisations and deaths in elderly long-term care facilities (LTCF) residents \citep{mazagatos2021effectiveness}. Moreover, these results showed a similar level of protection against asymptomatic and symptomatic infections among fully vaccinated LTCF residents. Another study performed in only one region of Spain, Navarra, showed that COVID-19 VE was moderate in preventing SARS-CoV-2 infection and was higher against symptomatic and hospitalised cases \citep{martinez2021effectiveness}. Therefore, these studies have already shown that vaccination has reduced hospitalization and lethality in Spain. However, the COVID-19 VE in reducing the COVID-19 transmission is still unclear, even though the study performed in Israel showed an association between COVID-19 incidence and vaccination \citep{haas2021impact}. Therefore, preventive measures such as social distance, masks wearing, hand washing, etc., which are very efficient in reducing the viral transmission \citep{wilder2020isolation} should be kept while SARS-CoV-2 is still circulating among us. 

In the present study, COVID-19 incidence and lethality rates have been studied at a regional analysis through segmented spatio-temporal models. This kind of macroscopic analysis does not allow quantifying VE, but enables to capture general trends in space and time. Particularly, the models fitted are able to detect change points in the vaccination levels that are reflected in the time series of either incidence or lethality rates. We have observed that increasing levels of vaccination display an association with reduced lethality rates. This has not been observed for incidence rates. Besides, the spatial component of the model has enabled us to determine incidence and lethality risks for the regions chosen for the analysis, whereas the consideration of a regionally-varying slope in the model has proven to be useful for capturing the differential behavior of some of the regions.

Finally, it is worth noting, again, that this type of ecological analysis has clear limitations, namely that it does not allow quantifying VE, but only to perform an exploratory (spatio-temporal) analysis of large-scale indicators. Nevertheless, these models could be helpful as an additional pandemic monitoring tool, especially if more covariates are added, or even if the possibility that the segmented covariate (in this case, the vaccination level) presents several change points is also incorporated. Regarding the inclusion of covariates, the consideration of binary variables that reflect how the restrictions and prevention measures have changed over time seems advisable to avoid confounding effects.



\section*{Funding}

Financial support was received from the Fundación Universidad Católica de Valencia San Vicente Mártir through Grant 2020-231-006.

\section*{Conflicts of interest}

The authors declare no conflict of interest in this study. 

\section*{Availability of data and material}

The data analyzed in this study is available upon reasonable request from the authors.

\section*{Code availability}

The R code created to carry out this study is available upon reasonable request from the authors.


\normalsize
\clearpage
\bibliographystyle{apalike}
\bibliography{bibliography}

\end{document}